\documentclass[aps,prb,a4paper,superscriptaddress,twocolumn,showpacs,amsmath,amssymb]{revtex4} 
\usepackage{graphicx,epsfig}
\usepackage[usenames]{color}

\allowdisplaybreaks
\begin{document}

\newcommand{\E}{\mathcal{E}}
\newcommand{\F}{\mathcal{F}}
\newcommand{\V}{\mathcal{V}}
\newcommand{\C}{\mathcal{C}}
\newcommand{\I}{\mathcal{I}}
\newcommand{\s}{\sigma}
\newcommand{\up}{\uparrow}
\newcommand{\dw}{\downarrow}
\newcommand{\h}{\mathcal{H}}
\newcommand{\g}{\mathcal{G}^{-1}_0}
\newcommand{\D}{\mathcal{D}}
\newcommand{\A}{\mathcal{A}}
\newcommand{\projs}{\hat{\mathcal{S}}_d}
\newcommand{\proj}{\hat{\mathcal{P}}_d}
\newcommand{\K}{\textbf{k}}
\newcommand{\Q}{\textbf{q}}
\newcommand{\T}{\tau_{\ast}}
\newcommand{\io}{i\omega_n}
\newcommand{\eps}{\varepsilon}
\newcommand{\+}{\dag}
\newcommand{\su}{\uparrow}
\newcommand{\giu}{\downarrow}
\newcommand{\0}[1]{\textbf{#1}}
\newcommand{\ca}{c^{\phantom{\dagger}}}
\newcommand{\cc}{c^\dagger}
\newcommand{\da}{d^{\phantom{\dagger}}}
\newcommand{\dc}{d^\dagger}
\newcommand{\fa}{f^{\phantom{\dagger}}}
\newcommand{\fc}{f^\dagger}
\newcommand{\ha}{h^{\phantom{\dagger}}}
\newcommand{\hc}{h^\dagger}
\newcommand{\be}{\begin{equation}}
\newcommand{\ee}{\end{equation}}
\newcommand{\bea}{\begin{eqnarray}}
\newcommand{\eea}{\end{eqnarray}}
\newcommand{\ba}{\begin{eqnarray*}}
\newcommand{\ea}{\end{eqnarray*}}
\newcommand{\dagga}{{\phantom{\dagger}}}
\newcommand{\bR}{\mathbf{R}}
\newcommand{\bQ}{\mathbf{Q}}
\newcommand{\bq}{\mathbf{q}}
\newcommand{\bqp}{\mathbf{q'}}
\newcommand{\bk}{\mathbf{k}}
\newcommand{\bh}{\mathbf{h}}
\newcommand{\bkp}{\mathbf{k'}}
\newcommand{\bp}{\mathbf{p}}
\newcommand{\bL}{\mathbf{L}}
\newcommand{\bRp}{\mathbf{R'}}
\newcommand{\bx}{\mathbf{x}}
\newcommand{\by}{\mathbf{y}}
\newcommand{\bz}{\mathbf{z}}
\newcommand{\br}{\mathbf{r}}
\newcommand{\Ima}{{\Im m}}
\newcommand{\Rea}{{\Re e}}
\newcommand{\Pj}[2]{|#1\rangle\langle #2|}
\newcommand{\ket}[1]{\vert#1\rangle}
\newcommand{\bra}[1]{\langle#1\vert}
\newcommand{\setof}[1]{\left\{#1\right\}}
\newcommand{\fract}[2]{\frac{\displaystyle #1}{\displaystyle #2}}
\newcommand{\Av}[2]{\langle #1|\,#2\,|#1\rangle}
\newcommand{\av}[1]{\langle #1 \rangle}
\newcommand{\Mel}[3]{\langle #1|#2\,|#3\rangle}
\newcommand{\Avs}[1]{\langle \,#1\,\rangle_0}
\newcommand{\eqn}[1]{(\ref{#1})}
\newcommand{\Tr}{\mathrm{Tr}}

\newcommand{\Vb}{\bar{\mathcal{V}}}
\newcommand{\Vd}{\Delta\mathcal{V}}
\def\P{P_{02}}
\newcommand{\Pb}{\bar{P}_{02}}
\newcommand{\Pd}{\Delta P_{02}}
\def\t{\theta_{02}}
\newcommand{\tb}{\bar{\theta}_{02}}
\newcommand{\td}{\Delta \theta_{02}}
\newcommand{\Rb}{\bar{R}}
\newcommand{\Rd}{\Delta R}

\title{Time-Dependent and Steady-State Gutzwiller approach for nonequilibrium \\transport in nanostructures}
\author{Nicola Lanat\`a}
\affiliation{Department of Physics and Astronomy, Rutgers University, Piscataway, New Jersey 08856-8019, USA} 
\author{Hugo U. R. Strand}
\affiliation{University of Gothenburg, SE-412 96 G\"oteborg, Sweden} 
\date{\today} 
\pacs{73.63.Kv, 73.63.-b, 71.27.+a}
\begin{abstract}
We extend the time-dependent Gutzwiller variational approach,
recently introduced by Schir\`o and Fabrizio,
Phys.~Rev.~Lett.~{\bf 105}~076401~(2010), to impurity problems.
Furthermore, we derive a consistent theory for the steady state, and show
its equivalence with the previously introduced nonequilibrium steady-state extension 
of the Gutzwiller approach.
The method is shown to be able to capture dissipation in the leads,
so that a steady state is reached after a sufficiently long relaxation time.
The time-dependent method is applied to the single orbital Anderson impurity model 
at half-filling, modeling a quantum dot coupled to two leads.
In these first exploratory calculations the Gutzwiller projector is limited 
to act only on the impurity.
The strengths and the limitations of this approximation are assessed via 
comparison with state of the art continuous time quantum Monte Carlo results.
Finally, we discuss how the method can be systematically improved
by extending the region of action of the Gutzwiller projector.
\end{abstract}

\maketitle

\section{Introduction}

The impressive advances in nanoelectronics has enabled contacting of very small
objects, such as quantum dots, molecular junctions and nanowires, with metallic 
electrodes, enabling very accurate measuraments of the source-drain current
counting individual electrons as they tunnel across the 
contact.~\cite{expcount1,expcount2,expcount3}
Because of the low dimensionality of the contact region correlations grow in strength
and may stabilize a local magnetic moment that influences electron 
tunneling.~\cite{GoldhaberGordon}

Due to the interplay between strong electron correlation and
out-of-equilibrium effects, the theoretical study of these systems is extremely
complicated.
Apart from many-body Keldysh perturbation theory,~\cite{Schoeller,pt-schmidt}
many innovative approaches have been explored,~\cite{NewJPhys.12.043042}
such as diagrammatic quantum Monte Carlo on the Keldysh 
contour,~\cite{diagrMC-schiro,diagrMC-werner1,diagrMC-werner2,diagrMC-muhlbacher,diagrMC-review}
field theory techniques,~\cite{field1,field2}
time dependent density matrix renormalization group,~\cite{td-dmrg-meisner,field2}
flow equation method,~\cite{flow-kehrein,flow-wang}
functional renormalization group,~\cite{fRG-jakobs,fRG-karrasch}
perturbative renormalization group,~\cite{pRG-rosch,pRG-chung,pRG-schoeller,pRG-Pletyukhov}
master equations,~\cite{masterEq-timm}
iterative path integral approaches,~\cite{it-pi-segal}
strong-coupling expansions,~\cite{scExp-mora1,scExp-vituschinsky,scExp-mora2,scExp-ratiani}
real time numerical renormalization group,~\cite{td-nRG-anders,td-nRG-schmitt}
scattering Bethe Ansatz,~\cite{mehta1}
and imaginary-time nonequilibrium quantum Monte Carlo.~\cite{itMC-Han1,itMC-Han2}

Unfortunately, most of these techniques are computationally very demanding, limiting their
application only to simple models.
For realistic cases, e.g., tunneling across a molecule or a transition metal atom --- where 
many orbitals participate to magnetism and affect conductance --- these methods become 
generally intractable.
In this general context a sufficiently simple approach to deal with realistic systems,
even one being less accurate than those previously mentioned, would be extremely useful.

With this motivation, an extension of the Gutzwiller variational
method~\cite{Gutzwiller1,Gutzwiller2} to nonequilibrium \emph{steady-state} 
transport was recently proposed in Ref.~\onlinecite{PhysRevB.82.195326},
and applied to the single-orbital Anderson impurity model at half-filling.
In parallel, a generalization of the \emph{time-dependent} Gutzwiller 
approach~\cite{lorenzana} was developed by
Schir\`o and Fabrizio~\cite{PhysRevLett.105.076401} on the basis of the 
Dirac-Frenkel variational principle,~\cite{dirac,frenkel,PhysRevB.83.165105}
and applied to study the quantum dynamics induced by an interaction quench
in the single band Hubbard model.

The first achievement of this paper is to generalize the time 
dependent Gutzwiller approach of 
Ref.~\onlinecite{PhysRevLett.105.076401} to impurity problems and derive a consistent 
theory for the steady state.
We find that dissipation occurs entirely within the leads 
as expected,~\cite{andrei-dissipation}
and the steady-state reached after a sufficiently long thermalization time
is the same as the one previously derived in Ref.~\onlinecite{PhysRevB.82.195326}.
We believe that our scheme represents a very interesting step toward the description
of steady state non-equilibrium problems. 
In fact, with this genuine variational principle
one can, in principle, always obtain more 
accurate results by extending the variational freedom of the wavefunction.

The second achievement of this paper is to investigate the reliability of 
the Gutzwiller variational ansatz in the case of a Gutzwiller projector acting only 
on the impurity (as was assumed in Ref.~\onlinecite{PhysRevB.82.195326}),  
which is the simplest conceivable Gutzwiller variational function
for an impurity model.
In this investigation we employ the single-orbital Anderson impurity model as a prototype,
and study two different ways to prepare the initial state before starting the 
time evolution: the ``interaction quench'' and the ``bias quench''.
We compare our calculations with the quantum Monte Carlo results 
of Ref.~\onlinecite{diagrMC-werner2}.
The comparison shows that our results are quantitatively predictive 
at short times.
At longer times, instead, the Gutzwiller dynamics can develop unphysical features, 
due to the oversimplified variational wavefunction.
We expect that the quality of the result would be considerably improved
if the Gutzwiller projector acted also on a portion of leads.
In particular, the resulting additional variational freedom would allow us to 
better account for the Kondo effect.

The outline of this paper is as follows.
In Sec.~\ref{approachsection} the time dependent Gutzwiller method is 
generalized to impurity models on the basis of the Dirac-Frenkel variational principle, 
and specialized to the single orbital Anderson impurity model at half-filling.
In Sec.~\ref{ststatesection} the variational approach for the steady state is 
derived on the same basis. Furthermore, several important conceptual differences with respect
to the theory at equilibrium are discussed.
In Sec.~\ref{numrelsection} the Gutzwiller variational results
for the interaction and the bias quench are presented. 
Both the transient and the long time dynamics are analyzed.
Finally, Sec.~\ref{conclusions} is devoted to conclusions.

\section{Time dependent Gutzwiller method for impurity models}\label{approachsection}

The Dirac-Frenkel variational principle,~\cite{dirac,frenkel} 
identifies the Schr\"odinger quantum dynamics of the system 
\be
\ket{\Psi(t)}=e^{-i\hat{\h} t}\ket{\Psi_i}
\ee
with the saddle point of the action
\bea
\mathcal{S}[\Psi(t)]&=&\int_{t_i}^{t_f} d\tau \mathcal{L}[\Psi(\tau)]\label{action}\\
\mathcal{L}[\Psi(\tau)]&=&\Av{\Psi(\tau)}{i\partial_\tau-\hat{\h}}\,,
\eea
searched in the set of all the possible time evolutions $\ket{\Psi(t)}$ in 
the Hilbert space --- with fixed boundaries at $t_i$ and $t_f$.
The advantage of the Dirac-Frenkel formulation of the dynamics
is that it allows us to build up a well founded variational approximation scheme. 
In fact, the saddle point of $\mathcal{S}$ can be searched in a proper subclass 
of the Hilbert space, chosen on the basis of physical motivations related 
with the specific system of interest.
An important example is the \emph{time-dependent Hartree-Fock method},
that can be derived as the saddle point of the Dirac-Frenkel action 
$\mathcal{S}$ assuming that $\ket{\Psi}$ spans the set of all Slater determinants.

As shown in Ref.~\onlinecite{PhysRevB.83.165105} by Schir\`o and Fabrizio, 
the Dirac-Frenkel principle provides a solid basis also
to the time-dependent Gutzwiller variational method, that was previously 
introduced by the same authors in Ref.~\onlinecite{PhysRevLett.105.076401} 
and used to study the quantum 
dynamics induced by an interaction quench in the single band Hubbard model.
The time-dependent Gutzwiller dynamics is expected to improve considerably the
time-dependent Hartree-Fock dynamics for interacting systems, 
as the Gutzwiller variational space enlarges the set of Slater determinants 
by means of the Gutzwiller projector, that is able to modify the weight
of local electronic configurations in accordance with the interaction.
In this section we show how to generalize the time-dependent Gutzwiller
scheme to study impurity problems.

\subsection{Gutzwiller dynamics of the single-orbital Anderson model}

For simplicity we employ the single-orbital Anderson model at half filling
--- modeling a quantum dot coupled to two leads ---
as a prototype for this scheme. 
The corresponding Hamiltonian is
\be
\hat{\mathcal{H}}=\hat{\mathcal{T}}+\hat{\mathcal{V}}+\hat{\mathcal{U}}\,,
\label{siam}
\ee
consisting of the following terms
\bea
&&\hat{\mathcal{T}}=\sum_{\alpha k\sigma}\epsilon_k\,\cc_{\alpha k\sigma}\ca_{\alpha k\sigma}
\label{hoppsiam}
\\
&&\hat{\mathcal{V}}=\sum_{\alpha k\sigma}\frac{V_k}{\sqrt{\Omega}}\,\cc_{\alpha k\sigma}\da_{\sigma}
\!+\!\text{H.c.}\label{hybrsiam}\\
&&\hat{\mathcal{U}}=\frac{U}{2}\left(\hat{n}_d-1\right)^2\,,\label{uint}
\eea
where $\cc_{\alpha k\sigma}$
creates a conduction electron on the left
$\alpha=-1$ or right $\alpha=1$ lead with quantum number $k$ and
spin $\sigma$ while $\dc_{\sigma}$ creates an electron in the dot with spin $\sigma$,
and $\Omega$ is the quantization volume of the system.

If the initial state $\ket{\Psi_i}$ is particle-hole symmetric 
then so is also $\ket{\Psi(t)}$ at any time. 
This allows us to restrict our attention to the particle-hole symmetric 
variational space.
We make the subsequent variational ansatz for the time-dependent 
wavefunction~\cite{PhysRevLett.105.076401}
\bea
\ket{\Psi}&=&\hat{\mathcal{P}}_G\ket{\Psi_0}=e^{-i\projs}\proj\,\ket{\Psi_0}\,,
\label{GWFch4ph}\\
\projs&=&\theta_{02}\hat{P}_{02}\label{improj}\\
\proj&=&l_{02}\hat{P}_{02}+\sqrt{2-l_{02}^2}\hat{P}_{1}
\eea
where $l_{02}$ and $\theta_{02}$ are real parameters,
\bea
\hat{P}_{02}\equiv \ket{\!\uparrow\downarrow}\bra{\uparrow\downarrow\!}\label{p02def}
\!+\!\ket{0}\bra{0}\,,\;
\hat{P}_{1}\equiv \ket{\!\uparrow}\bra{\uparrow\!}
\!+\!\ket{\!\downarrow}\bra{\downarrow\!}
\eea
are impurity projection operators, and $\ket{\Psi_0}$ is a Slater determinant.
It can be easily verified that ${P}_{02}\equiv\Av{\Psi}{\hat{P}_{02}}=l_{02}^2/2$,
so that ${P}_{02}$ and $\theta_{02}$ can be used as variational parameters.
Furthermore, it can be easily shown that the wavefunction $\ket{\Psi}$ defined above
satisfies the Gutzwiller conditions
\bea
\Av{\Psi_0}{\mathcal{P}_G^\dagger\mathcal{P}^\dagga_G}\!&=&\!1\label{gc1}\\
\Av{\Psi_0}{\mathcal{P}_G^\dagger\mathcal{P}^\dagga_G\,\dc_\sigma\da_\sigma}\!
&=&\!\frac{1}{2}\label{gc2}
\eea
for all possible values of $0\leq{P}_{02}\leq 1$ and $\theta_{02}$.
Notice that $P_{02}$ is just twice the expectation value of the double-occupancy 
of the dot.

Using Eqs.~\eqref{gc1} and \eqref{gc2} one finds that the average energy $\E$ 
with respect to the Gutzwiller variational wavefunction~[Eq.~\eqref{GWFch4ph}] 
is given by
\bea
\E[P_{02},\theta_{02},\Psi_0]&=&\Av{\Psi}{\hat{\h}}\nonumber\\
&=&\Av{\Psi_0}{\hat{\h}^{R}_0}+U{P}_{02}/2\,,
\label{varanszph}
\eea
where
\bea
\hat{\mathcal{H}}^{R}_0&=&\hat{\mathcal{T}}+R\hat{\mathcal{V}}\label{hrenph}\\
R&=&2\sqrt{{P}_{02}(1-{P}_{02})}\cos(\theta_{02})\label{rph}\,.
\eea
Furthermore, from Eqs.~\eqref{gc1} and \eqref{gc2} we obtain the following
analytical expression for the Lagrangian~\cite{PhysRevB.83.165105}
\bea
\mathcal{L}[P_{02},\theta_{02},\Psi_0]&=&P_{02}\partial_t\theta_{02}+
\langle\Psi_0|i\partial_t\Psi_0\rangle\nonumber\\
&-&\E[P_{02},\theta_{02},\Psi_0]\,.\label{lagandph}
\eea
The corresponding equations of motion are
\bea
\frac{\partial{P}_{02}}{\partial t}&=&-\frac{\partial \E}{\partial\theta_{02}}\label{dyn1ph}\\
\frac{\partial{\theta}_{02}}{\partial t}&=&\frac{\partial \E}{\partial{P}_{02}}\label{dyn2ph}\\
\frac{\partial\ket{\Psi_0}}{\partial t}&=&-i\hat{\mathcal{H}}^{R}_0\ket{\Psi_0}\label{dyn3ph}\,.
\eea

Let us now rewrite Eqs.~\eqref{dyn1ph}-\eqref{dyn3ph} in a soluble form.
Substituting Eqs.~\eqref{varanszph} and \eqref{hrenph} in 
Eqs.~\eqref{dyn1ph} and \eqref{dyn2ph} gives
\bea
\frac{\partial{P}_{02}}{\partial t}\!\!&=&\!\!
-\mathcal{V}(t)\frac{\partial{R}}{\partial \theta_{02}}\label{dyn1bisph}\\
\frac{\partial{\theta}_{02}}{\partial t}\!\!&=&\!\!
\frac{U}{2}+\mathcal{V}(t)\frac{\partial{R}}{\partial P_{02}}
\label{dyn2bisph}
\eea
where $\mathcal{V}(t)=\Av{\Psi_0(t)}{\hat{\mathcal{V}}}$.
The time evolution of $\ket{\Psi_0}$, see Eq.~\eqref{dyn3ph}, 
can be formulated in the Heisenberg picture as follows:
\bea
\label{d1}
&&\frac{\partial\sqrt{\Omega}\av{\cc_{\alpha k\sigma}\da_{\sigma}}}{\partial t}=
i\left\{
\epsilon_k\sqrt{\Omega}\av{\cc_{\alpha k\sigma}\da_{\sigma}}+
\right.\nonumber\\&&\;\left.
RV\av{\dc_{\sigma}\da_{\sigma}}
-\frac{1}{\Omega}{\sum_{\alpha'k'}}
RV\,\Omega\av{\cc_{\alpha k\sigma}\ca_{\alpha' k'\sigma}}
\right\}\\
\label{d2}
&&\frac{\partial\Omega\av{\cc_{\alpha k\sigma}\ca_{\alpha' k'\sigma}}}{\partial t}=
i\left\{
(\epsilon_k-\epsilon_{k'})\,\Omega\av{\cc_{\alpha k\sigma}\ca_{\alpha' k'\sigma}}+
\right.\nonumber\\&&\;\,\left.
RV\left(\sqrt{\Omega}\av{\da_{\sigma}\cc_{\alpha' k'\sigma}}-
\sqrt{\Omega}\av{\cc_{\alpha k\sigma}\da_{\sigma}}\right)
\right\}
\,,
\eea
where, for simplicity, a constant $V_k\equiv V$ has been assumed, 
and the averages are taken with respect to $\ket{\Psi_0(t)}$.
Notice that the occupancy of the dot $\av{\dc_{\sigma}\da_{\sigma}}$ 
is $1/2$ due to particle-hole symmetry.

The system defined by Eqs.~\eqref{dyn1bisph}-\eqref{d2} constitutes a set of coupled 
first order differential equations, which can be readily solved numerically.
From the time evolution of the parameters that appear in Eqs.~\eqref{dyn1bisph}-\eqref{d2}
is possible to calculate the expectation value of any desired observable.
In particular, the expectation value of the current is obtained as
\bea
&I_\alpha[\Psi] & =  -i \frac{V}{\sqrt{\Omega}} \sum_{ k \sigma} \left(
\Av{\Psi}{\dc_\sigma\ca_{\alpha k\sigma}} - \text{c.c.} \right)
\nonumber \\ && =  
-2 \frac{RV}{\Omega} \sum_{ k \sigma}  
\text{Im}\left[\sqrt{\Omega}\av{\cc_{\alpha k\sigma}\da_{\sigma}}\right]\,.
\label{tdGcurr}
\eea
In the particle-hole symmetric case the left and right currents are equal 
in magnitude and the lead index $\alpha$ can be dropped, $I \equiv |I_\alpha|$.

\section{Gutzwiller theory for the steady state}\label{ststatesection}

Let us consider a quantum system whose dynamics is generated by the 
action [Eq.~\eqref{action}] in a Hilbert space $\I$.
The condition of stationarity for a state $\ket{\Psi}$ is that
\be
\left.\frac{\delta\mathcal{L}(\Psi,\partial_t\Psi)}{\delta\Psi}\right|_{\partial_t\Psi=0}\!=0
\quad\forall\,\ket{\delta\Psi}\!\perp\!\ket{\Psi}\in\I\,.
\label{stat_action}
\ee
In this section we will derive the Gutzwiller theory for the steady state
starting from the general condition 
[Eq.~\eqref{stat_action}] applied to the Dirac-Frenkel dynamics
in the Gutzwiller variational space.

For clarity we will focus on the single orbital Anderson impurity model at half-filling
and \emph{infinite} leads. 
This will facilitate the subsequent discussion of the most general case.
Furthermore, we will assume --- without any loss of generality~\cite{Hewson} --- 
a one-dimensional representation of the Hamiltonian of the leads [Eq.~\eqref{hoppsiam}]
\be
\hat{\mathcal{T}}=\sum_{\alpha\sigma}\sum_{RR'}
t_{RR'}\,\cc_{\alpha R\sigma}\ca_{\alpha R'\sigma}\,,
\ee
where the site labels $R$ and $R'$ run over all integer numbers from $1$ to $\infty$.

\subsection{Infinite systems and inequivalent representations}\label{brokensym}

Before proceeding with the derivation of the Gutzwiller scheme,
we need a preliminary discussion of the problem of the steady state in 
relation with the general structure of infinite 
fermionic systems.

Let us define the $C^*$-algebra~\cite{Sewell,BratteliRobinson} $\mathcal{A}$ generated
by the \emph{local} fermionic ladder operators $\cc_{\alpha R\sigma}$ ---
satisfying the canonical anticommutation relations.

We associate to a generic state $\ket{\Psi}$ of the Anderson system [Eq.~\eqref{siam}]
its corresponding \emph{sector} $\I[\Psi]$, defined as the Hilbert space spanned by all 
the states $\hat{A}\ket{\Psi}$, with $\hat{A}\in\mathcal{A}$.
The linear space $\I[\Psi]$ is the basis of a representation of $\mathcal{A}$, i.e., 
\be
\ket{\phi}\in\I[\Psi]\Rightarrow\hat{A}\ket{\phi}\in\I[\Psi]
\quad\forall\hat{A}\in\mathcal{A}\,.
\label{repr}
\ee

The physical meaning of Eq.~\eqref{repr} is that $\I[\Psi]$ represents a class 
of ``macroscopically equivalent'' states, that differs from $\ket{\Psi}$ only by
local modifications.
In this generalized sense, a sector can be regarded as a 
\emph{phase} of the system.~\cite{Sewell}

Note that, in general, it is not guaranteed that the sectors of two different states coincide.
In fact, the Von Neumann theorem does not apply to infinite systems,~\cite{Sewell}
and the observables generally admit a big variety of inequivalent 
representations.

Let us define the initial state $\ket{\Psi_{\text{in}}}$ of 
two unconnected leads prepared at chemical potentials $\mu_\alpha=\alpha\Phi/2$,
where $\Phi$ corresponds to the applied bias.
Once the tunneling region [Eq.~\eqref{hybrsiam}] is connected to the leads, 
the initial state $\ket{\Psi_{\text{in}}}$ evolves according to the  Schr\"odinger equation
\be
i\partial_t\ket{\Psi(t)}=\hat{\h}\ket{\Psi(t)}\,,
\ee
with $\ket{\Psi(t)}\in\mathcal{I}[\Psi_{\text{in}}]$ for any \emph{finite} time $t$.
A very important point for our subsequent discussion is that 
the steady state $\ket{\Psi_S}$, that is reached after an \emph{infinite} time,
does not belong to $\mathcal{I}[\Psi_{\text{in}}]$.

Let us prove our statement. 
It is clear that all the vectors that belong to the same sector share the 
same \emph{average} left-current, which is defined as follows:
\be
\bar{I}[\Psi]=
\lim_{\Lambda\uparrow}\frac{1}{|\Lambda|}\Av{\Psi}{\sum_\alpha\sum_{R\in\Lambda}\hat{I}_{R\alpha}}\,,
\label{iav}
\ee
where the symbol $\Lambda\!\!\uparrow$ denotes a sequence of bounded subregions
$\Lambda$ of the system that increase to \emph{infinity}, 
$|\Lambda|$ is the number of sites in $\Lambda$, 
\be
\hat{I}_{R\alpha}=i[\hat{\h},\hat{N}_{R\alpha}]\,,
\ee
and $\hat{N}_{R\alpha}$ represents the total number of electrons to the left of 
the site identified by $R$ and $\alpha$.
In fact, the limit in Eq.~\eqref{iav} is not affected by the contribution of any 
local operator $\hat{A}$, i.e., 
\be
\bar{I}[\Psi]=\bar{I}[\hat{A}\Psi]\quad\forall \hat{A}\in\A\,.
\label{barI}
\ee

The meaning of Eq.~\eqref{barI} is that two macroscopically equivalent states
share the same average current.
In this generalized sense,~\cite{Sewell} the average current can be regarded 
as an ``order parameter'', that enables us to distinguish different sectors (phases) 
of states.

This simple observation makes it easy to prove the statement that 
$\ket{\Psi_S}$ does not belong to $\mathcal{I}[\Psi_{\text{in}}]$.
In fact, after an infinite time, all
expectation values of the local current in Eq.~\eqref{iav} are non-zero, 
so that 
\be
\bar{I}[\Psi_S]\neq\bar{I}[\Psi_{\text{in}}]=0\,.
\ee
The thesis follows by contradiction from the fact proven above 
that if $\ket{\Psi_{\text{in}}}$ and $\ket{\Psi_S}$ would belong to the same 
sector then they would share the same average current.

In particular, this observation implies that the limit
\be
\ket{\Psi_S}=\lim_{t\rightarrow\infty}\ket{\Psi(t)}
\ee
can \emph{not} be interpreted as a limit in the norm induced by the scalar 
product in a Hilbert space, but is well defined only in the following ``weak'' sense
\be
\Av{\Psi_S}{\hat{A}}=
\lim_{t\rightarrow\infty}\Av{\Psi(t)}{\hat{A}}\quad\forall \hat{A}\in\A\,.
\ee

Furthermore, the same argument shows that steady states of Hamiltonians that differ only locally
--- e.g., at the impurity --- belong, in general, to inequivalent representations of the 
observables. 
In fact, the steady state average current depends on the local interaction at the impurity.

Notice that the argument above does not apply to the equilibrium case, as
the average current is zero independently on the junction at zero bias.
In fact, in this case (and only in this case), $\ket{\Psi_{\text{in}}}$ is the ground state
of $\hat{\mathcal{T}}$, so that the action to connect the tunneling region
to the leads corresponds to perturb locally the ground state of $\hat{\mathcal{T}}$
with 
\be
\delta\hat{\h}\equiv\hat{\mathcal{V}}+\hat{\mathcal{U}}\,,
\ee
see Eqs.~\eqref{siam}-\eqref{uint}.
This implies that $\ket{\Psi_{\text{S}}}$
is the ground state of $\hat{\h}$, which
belongs to the same sector of $\ket{\Psi_{\text{in}}}$.
This is ensured by well established  general results concerning the 
``stability of thermal sectors under local perturbations'', 
see Ref.~\onlinecite{BratteliRobinson}.

In conclusion, the steady states of biased systems with different junctions belong 
to disjoint representations of the observables.  
This is a fundamental difference 
between equilibrium and out-of-equilibrium problems, and will be central
in our subsequent discussion of the Gutzwiller theory for 
the nonequilibrium steady state.

\subsection{Steady state theory for the single orbital Anderson impurity model at half-filling}
\label{ssSIAM}

In this section we show how the Gutzwiller steady-state theory
for the single orbital Anderson impurity model at 
half-filling~\cite{PhysRevB.82.195326} can be derived
by means of the general scheme introduced above.

From Eq.~\eqref{lagandph} we have that the stationarity condition 
[Eq.~\eqref{stat_action}] is equivalent to
\bea
\frac{\delta\E(\Psi_0,P_{02},\theta_{02})}{\delta\Psi_0}&=&0
\label{st_act_siam-ph1}\\
\frac{\partial\E(\Psi_0,P_{02},\theta_{02})}{\partial P_{02}}&=&0
\label{st_act_siam-ph2}\\
\frac{\partial\E(\Psi_0,P_{02},\theta_{02})}{\partial\theta_{02}}&=&0
\label{st_act_siam-ph3}\,,
\eea
where $\E$ is given by Eq.~\eqref{varanszph} and $\ket{\delta{\Psi_0}}\in\I[\Psi_0]$ 
is orthogonal to $\ket{\Psi_0}$.

Let us assume that the Gutzwiller parameters ${P}_{02}$ and $\theta_{02}$ converge to
proper steady state values after a relaxation time $T$ due to 
dissipation within the leads,~\cite{andrei-dissipation} 
see Sec.~\ref{numrelsection}.
This implies that for $t>T$ only the Slater determinant $\ket{\Psi_0}$ 
evolves in time, and its dynamics is induced by the effective steady-state Lagrangian
\be
\mathcal{L}_0[\Psi_0]=
\Av{\Psi_0}{i\partial_t-\hat{\h}^{R}_0}\,,
\ee
where $R$ is obtained from  $P_{02}$ and $\theta_{02}$ using Eq.~\eqref{rph}.
Equivalently, for $t>T$
\be
\ket{\Psi_0(t)}=e^{-i\hat{\h}^{R}_0 (t-T)}\ket{\Psi_0(T)}\,.
\ee
From (i) the uniqueness of the steady state and (ii) the observation that 
$\ket{\Psi_0(T)}$ and $\ket{\Psi_0(0)}$ differ only locally 
we obtain that, after an infinite time, the Slater determinant is given by
\be
\ket{\Psi^{R}_0}=\lim_{t\rightarrow\infty}e^{-i\hat{\h}^{R}_0 t}\ket{\Psi_0(0)}\,,
\label{uncorrsteady}
\ee
which is simply the steady state of $\hat{\h}^{R}_0$.
The above argument reproduces the \emph{variational ansatz} 
for the steady state used in Ref.~\onlinecite{PhysRevB.82.195326} 
from a very clear perspective.  In order to not evolve in time with 
Eqs.~\eqref{dyn1ph}-\eqref{dyn3ph}, the Slater determinant $\ket{\Psi_0}$
of a variational state $\ket{\Psi}$ of the form [Eq.~\eqref{GWFch4ph}] 
is necessarily the steady state $\ket{{\Psi}^{R}_0}$ of $\hat{\h}^{R}_0$, 
where ${R}$ is obtained from  ${P}_{02}$ and ${\theta}_{02}$ using 
Eq.~\eqref{rph}.

This consideration facilitates the search for the steady state.
In fact, we can restrict the search of the state that satisfy the 
stationarity conditions [Eqs.~\eqref{st_act_siam-ph1}-\eqref{st_act_siam-ph3}] 
to the variational subset of ``candidate'' Gutzwiller steady states 
of the form defined above,
\be
\ket{\Psi}=\hat{\mathcal{P}}_G\ket{\Psi_0^R}\,,
\label{formsteadystate}
\ee
each one belonging to a distinct representation of the observables.
In our case, this amounts to search for the solution of 
Eqs.~\eqref{st_act_siam-ph1} and \eqref{st_act_siam-ph2} only, 
as Eq.~\eqref{st_act_siam-ph3} is automatically satisfied by $\ket{\Psi^{R}_0}$.
From the stationarity conditions [Eqs.~\eqref{st_act_siam-ph1} and \eqref{st_act_siam-ph2})] 
we obtain that
\bea
\frac{U}{2}+\Av{\Psi^R_0}{\hat{\mathcal{V}}}\frac{\partial R}{\partial P_{02}}&=&0\label{s1}\\
\Av{\Psi^R_0}{\hat{\mathcal{V}}}\frac{\partial R}{\partial \theta_{02}}&=&0\label{s2}\,,
\eea
which ensure that ${P}_{02}$ and ${\theta}_{02}$ do not evolve in time,
see Eqs.~\eqref{dyn1bisph} and \eqref{dyn2bisph}).
Notice that $\Av{\Psi^R_0}{\hat{\mathcal{V}}}$ can be readily calculated by means of 
the Keldysh formalism~\cite{Schoeller} or scattering 
theory,~\cite{Hanscatt,annalshan} 
as $\hat{\h}^{R}_0$ is a quadratic Hamiltonian, see section \ref{steadystatesection}.

Eq.~\eqref{s2} implies that the Gutzwiller projector of the Gutzwiller
steady state is real, as it implies that $\theta_{02}$ is a multiple of $\pi$,
see Eq.~\eqref{improj}.
Eq.~\eqref{s1} allows us to calculate the steady state value of $P_{02}$, as
$R$ is a known function of $P_{02}$, see Eq.~\eqref{rph}.

In conclusion, we have shown that if a Gutzwiller steady-state exists 
then it is determined by 
Eqs.~\eqref{formsteadystate}-\eqref{s2}.

Note that, from the general point of view of Sec.~\ref{brokensym}, 
Eqs.~\eqref{st_act_siam-ph1}-\eqref{st_act_siam-ph3}
establish the condition of stationarity of a given variational state $\ket{\Psi}$
in its \emph{own} sector $\I[\Psi]=\I[\Psi_0^R]$,
which is indeed, contrarily to the equilibrium case, 
a \emph{state-dependent} variational space.

\subsection{Steady state theory for multi-orbital Anderson Models}\label{sec:generalcase}

In this section we generalize the Gutzwiller steady-state theory to 
a general impurity model representing two infinite leads connected through a generic
multi-orbital junction
\be
\hat{\h}=\hat{\mathcal{T}}+\hat{\mathcal{V}}+\hat{\mathcal{U}}\,,
\label{general}
\ee
where $\hat{\mathcal{U}}$ is the Hamiltonian of the junction, and
\bea
&&\hat{\mathcal{T}}=\sum_{\alpha k\sigma}\epsilon_k\,\cc_{\alpha k\sigma}\ca_{\alpha k\sigma}
\\
&&\hat{\mathcal{V}}=\sum_{\alpha k\sigma}\sum_n\frac{V^n_{k\alpha}}{\sqrt{\Omega}}
\,\dc_{n\sigma}\ca_{\alpha k\sigma}
\!+\!\text{H.c.}\,,
\eea
where $\cc_{\alpha k\sigma}$ creates a conduction electron on the left
$\alpha=-1$ or right $\alpha=1$ lead with quantum number $k$ and
spin $\sigma$, $\dc_{n\sigma}$ creates an electron in the junction with 
quantum label $n$ and spin $\sigma$,
and $\Omega$ is the quantization volume of the system.

We employ the general $\phi$-matrix formalism, see Ref.~\onlinecite{lanata} 
and references therein, assuming, for simplicity, a spin rotationally invariant 
Gutzwiller variational state. 
Let us introduce the so-called 
\emph{natural-basis}~\cite{dimerboss} operators $\fa_{n\sigma}$, i.e., the ladder 
operators such that  
\be
\Av{\Psi_0}{f^\dagger_{n\sigma}f^\dagga_{m\sigma}} = \delta_{nm}\,n^0_{n} 
\quad\forall\,n,m
\label{C-avch1}
\ee
where $0\leq n^0_{\alpha}\leq 1$ are the eigenvalues of the local density matrix
\be
\Av{\Psi_0}{d^\dagger_{n\sigma}d^\dagga_{m\sigma}}
\equiv\rho^0_{nm}\,.
\label{ldmcbasis}
\ee
Furthermore, we introduce the matrix representation of the operators $f^\dagga_{n\sigma}$
\be
\left(F^\dagga_{n\sigma}\right)_{ij}=
\langle \Gamma_i|f^\dagga_{n\sigma}|\Gamma_j\rangle\,,
\ee
where $\ket{\Gamma_i}$ are many-body Fock states expressed in the 
$f_{n\sigma}$-basis.

It can be shown that the Gutzwiller variational function 
is parametrized by $\phi$ and $\ket{\Psi_0}$ satisfying the 
Gutzwiller constraints
\bea
\Tr\left(\phi^\dagger\phi\right) \!\!&=&\!\! 1\,,\label{cc1}\\
\Tr\left(\phi^\dagger\phi\,F^\dagger_{n\sigma}F^\dagga_{m\sigma}\right) 
\!\!&=&\!\!n^0_{n}\delta_{nm}\,,
\label{cc2}
\eea
and the variational energy  is given by 
\bea
\E[\phi,\Psi_0]&=&\Av{\Psi_0}{\hat{\h}^{R}_0}
+\Tr\left(\phi^\dagger\, U\,\phi\right)\,,
\label{variationalenergych1}
\eea
where
\bea
\hat{\h}^{R}_0&=&\hat{\mathcal{T}}+\hat{\mathcal{V}}_R
\label{ttilde}\\
\hat{\mathcal{V}}_R&=&\sum_{\alpha k\sigma}\sum_n
\frac{\sum_mR_{mn}V^n_{k\alpha}}{\sqrt{\Omega}}
\,\fc_{n\sigma}\ca_{\alpha k\sigma}
\label{eq:RenormalizedHopping}
\\
R_{nm} &=& \frac{\Tr(\phi^{\dagger}\,F^\dagger_{n\sigma}\,\phi\,F^\dagga_{m\sigma})}
{\sqrt{n^0_{m}(1-n^0_{m})}}
\\
U_{ij}&=&\langle \Gamma_i|\,\hat{\mathcal{U}}\,|\Gamma_j\rangle
\,.\label{Z-newch1}
\eea

It can be easily shown that the Dirac-Frenkel Lagrangian is given by
\bea
\mathcal{L}[\phi,\Psi_0,\lambda]\!\!&=&\!\!
\langle\Psi_0|i\partial_t\Psi_0\rangle
-\E[\phi,\Psi_0]\label{totallagrgen}
\nonumber\\
\!\!&+&\!\!\Av{\Psi_0}{\hat{\mathcal{P}}_G(\partial_t\hat{\mathcal{P}}_G)}
+C[\lambda;\phi,\Psi_0]\,,
\eea
where $C$ is a Lagrange multiplier term that 
ensures that the Gutzwiller constraints 
-- which can be regarded as holonomic constraints --
are satisfied at any time.

Let us assume to have solved the corresponding Lagrange equations.
Similarly to the case of the single orbital Anderson impurity model,
the matrix $\phi$, the variational density matrix $n^0$ and the 
Lagrange multipliers $\lambda$ converge to
proper steady state values after a transient time $T$. 
This implies that for $t>T$ only the Slater determinant $\ket{\Psi_0}$ 
evolves in time by means of the dynamics induced by the 
effective steady-state Lagrangian
\be
\mathcal{L}_0[\Psi_0]=
\Av{\Psi_0}{i\partial_t-[\hat{\h}^{R}_0+\!\sum_{nm}\lambda_{nm}\!\sum_\sigma\fc_{n\sigma}\fa_{m\sigma}]}\,,
\ee
where $R$ is obtained from  $\phi$ and $n^0$ using 
Eq.~\eqref{Z-newch1}.
In order to not evolve in time, $\ket{\Psi_0}$ is necessarily the 
\emph{steady state} of 
\be
\hat{\h}_0^*[R,\lambda]\equiv\hat{\h}^{R}_0+\sum_{nm}{\lambda}_{nm}\sum_\sigma\fc_{n\sigma}\fa_{m\sigma}\,,
\label{hrengen}
\ee 
where ${R}$ is obtained from $\phi$ and $n^0$ using 
Eq.~\eqref{Z-newch1} and $\lambda$ ensures that 
$\ket{\Psi_0}$ satisfies Eq.~\eqref{C-avch1}.

From Eq.~\eqref{totallagrgen} we deduce that the stationarity condition
[Eq.~\eqref{stat_action}] is equivalent to
\be
\delta\E(\Psi_0,\phi)=0\label{vargeneric}
\ee
for all the variations of $\phi$ and $\ket{\delta\Psi_0}\in\I[\Psi_0]$ 
such that $\ket{\delta\Psi_0}\perp\ket{\Psi_0}$ and 
the ``holonomic'' constraints [Eqs.~\eqref{cc1} and \eqref{cc2}] are satisfied.
Similarly to the case of the single orbital Anderson impurity model, the argument
above facilitates the search of the steady state,
as the solution of Eq.~\eqref{vargeneric} can be searched in
the restricted variational subspace for the nonequilibrium steady state.

We point out that Eq.~\eqref{vargeneric} 
establishes the condition of stationarity of a variational state
$\ket{\Psi}$ in $\mathcal{I}[\Psi]$, that is a 
\emph{state-dependent} variational space.

Notice that, in principle, the general scheme described above
allows us to study any impurity system with an \emph{arbitrary level of accuracy}.
Let us consider, for instance, the single orbital Anderson impurity model.
One can treat the impurity and a portion of leads as a multi-orbital junction
and study the system with the general method described above.
The resulting increased variational freedom would 
take into account also the local magnetic correlations between the impurity and the 
surrounding lead electrons.
The so obtained approximated steady state is anticipated to converge 
to the exact steady state upon increasing the size of the portion of leads included in 
the projected region. 
In fact, one can presumably neglect any direct influence that
the interaction in the scattering region may give to the leads sufficiently far
from the impurity because of the screening effect.

Finally, we observe that, from the technical point of view, the complexity of the 
Gutzwiller problem is the same as in the equilibrium case.
The only difference is that the calculation of the 
ground state of the quadratic Hamiltonian $\hat{\h}_0^*[R,\lambda]$, see
Eq.~\eqref{hrengen}, is replaced with the calculation of its steady state.
It follows that the technical limit of the approach at finite bias is 
the same as in the equilibrium case: the number of variational parameters 
scales exponentially  with the number of orbitals of the junction.

\subsection{Possible connection with the Hershfield effective equilibrium theory}\label{generalsection}

In his seminal work~\cite{Her.} Hershfield proposed an alternative description of 
the steady state in terms of an effective
equilibrium theory defined by a modified Hamiltonian of the form
\be
\hat{\h}_\Phi\equiv\hat{\h}+\Phi\hat{\mathcal{Y}}\,.
\ee
It can be proven~\cite{annalshan} that the nonequilibrium steady state
can be expressed in a unique way in the Boltzmann form
\be
\rho_\Phi\propto\exp(-\beta\hat{\h}_\Phi)\,,
\ee
where the so called ``bias operator'' $\hat{\mathcal{Y}}$ encodes the dependency
on the bias. 
In the latest years the interest in this alternative formulation 
has grown considerably.
In fact, beyond its purely conceptual relevance in the general context 
of nonequilibrium statistical 
mechanics,~\cite{matsui-ogata,araky-moriya}
several promising numerical techniques have successfully implemented this 
scheme.~\cite{td-nRG-anders,td-nRG-schmitt,itMC-Han1,itMC-Han2}

In this section we discuss a general aspect of our variational scheme 
in relation with the effective equilibrium Hershfield framework.

Let us consider a Gutzwiller wavefunction of the form
\be
\ket{\Psi_\Lambda}=\hat{\mathcal{P}}^\Lambda_G\ket{\Psi_0}\label{ansatzLambda}\,,
\ee
where $\Lambda$ is a subregion of the system that contains the impurity, and
$\hat{\mathcal{P}}^\Lambda_G$ is the most general operator generated 
by any algebraic combination of local fermionic ladder operators $\cc_{\alpha R\sigma}$
and $\ca_{\alpha R\sigma}$ with $R\in\Lambda$.

As discussed in the previous section, we do not expect that
the interaction affects directly the electrons  
sufficiently far from the scattering region --- because of the screening effect.
This observation suggests that the Gutzwiller steady state 
variationally determined from the ansatz [Eq.~\eqref{ansatzLambda}] 
\be
\ket{\Psi_{\Lambda S}}\equiv\hat{\mathcal{P}}^{\Lambda S}_{G}\ket{\Psi^{\Lambda S}_0}
\label{varappr}
\ee
is a good approximation of the exact steady state $\ket{\Psi_S}$ if $\Lambda$ is 
sufficiently large.
In other words, we expect that the Gutzwiller projector 
$\hat{\mathcal{P}}^{\Lambda S}_{G}$ produces important variational corrections by increasing
$\Lambda$ only up to a certain finite region $\bar{\Lambda}$
(localized around the correlated impurity), 
and that further increase the variational freedom can provide only ``marginal'' 
corrections.

The above conjecture is formulated mathematically by the following equations:
\bea
\lim_{|\Lambda|\uparrow}\Av{\Psi_{\Lambda S}}{\hat{A}}&=&
\Av{\Psi_S}{\hat{A}}\quad\forall\hat{A}\in\A\label{existlim}\\
\lim_{|\Lambda|\uparrow}\Av{\Psi^{\Lambda S}_0}{\hat{A}}&\equiv&
\Av{\Psi^0_S}{\hat{A}}\quad\forall\hat{A}\in\A\label{lim1}\\
\mathcal{I}[\Psi^0_S]&=&\mathcal{I}[\Psi_S]
\label{lim2}\,.
\eea

The meaning of Eq.~\eqref{existlim} is that the series of Gutzwiller approximations 
[Eq.~\eqref{varappr}] ``approaches'' the exact steady state 
in the limit of infinite $\Lambda$.
Eq.~\eqref{lim1} expresses the existence of the weak limit of the
corresponding series of Slater determinants $\ket{\Psi^{\Lambda S}_0}$.
Finally, Eq.~\eqref{lim2} asserts that the exact 
correlated steady state $\ket{\Psi_S}$ belongs to the same sector of a proper uncorrelated Slater 
determinant $\ket{\Psi^0_S}$, i.e., that $\ket{\Psi_S}$ and $\ket{\Psi^0_S}$ differ
only by local modifications, see Sec.~\ref{brokensym}. 
If proven, this would be an interesting observation in relation with 
the general theory of infinite systems.~\footnote{We thank Giovanni Morchio for pointing this out.}

From the point of view of the Hershfield theory, the verification of our conjecture would
suggest that, given a general correlated impurity problem, 
the bias operator $\hat{\mathcal{Y}}$ can be expressed as 
\be 
\hat{\mathcal{Y}}\simeq\hat{\mathcal{Y}}_0+\delta \hat{\mathcal{Y}}\,,
\label{nozieres-bias}
\ee
where $\delta \hat{\mathcal{Y}}$ is a 
\emph{local} operator, and $\hat{\mathcal{Y}}_0$ is a proper ``long-range''
quadratic operator that depends on the Hamiltonian of the system $\hat{\h}$.

Note that to a local difference between two Hamiltonians 
$\hat{\h}_A$ and $\hat{\h}_B$ corresponds a ``macroscopic'' 
difference between the corresponding effective long-range potential 
$\hat{\mathcal{Y}}^A_0$ and $\hat{\mathcal{Y}}^B_0$.
This is consistent with our observation that 
the sector of steady states of Hamiltonians that differ only locally
belong to different representations of the observables, i.e.,
they are macroscopically different, see Sec.\ref{brokensym}.

The physical meaning of Eq.~\eqref{nozieres-bias} is in line
with the concept of nonequilibrium 
Nozi\'eres Fermi liquid theory~\cite{Nozieres-Journal-of-Low-Temperature-Physics}
in the form proposed in Ref.~\onlinecite{PhysRevB.82.195326}:
it should be possible to represent the 
properties of the system in terms of weakly interacting quasi-particles which,
by continuity with the non-interacting case, should be regarded as scattering states
of a properly renormalized quadratic Hamiltonian.

\section{Numerical results}\label{numrelsection}

In this section we investigate the reliability of the variational ansatz
[Eq.~\eqref{GWFch4ph}] employing the single orbital Anderson impurity model at 
half-filling as a prototype.

We define the hybridization function as
\be
\Delta(\epsilon) = \sum_k\frac{V_k^2}{\Omega}\frac{1}{\epsilon-\epsilon_k+i0^+}
\ee
and the hybridization width as
\be
\Gamma(\epsilon)\equiv\pi\sum_k\frac{V_k^2}{\Omega}\,\delta(\epsilon-\epsilon_k)\,.
\ee

The half-bandwidth $W$ of the leads will be used as the unit of energy.

\subsection{The steady state}\label{steadystatesection}

In Sec.~\ref{ssSIAM} we derived the Gutzwiller nonequilibrium 
steady state for the single orbital Anderson impurity model.
We also demonstrated that in the steady state the phase $\tb$ is zero, 
see Eq.~\eqref{s2}, and that $\Pb$ is determined by the condition
\be
\frac{U}{2} + \Av{\Psi^{\Rb}_0}{\hat{\mathcal{V}}} \frac{1-2\Pb}{\sqrt{\Pb(1-\Pb)}} = 0
\ee
where
\be
\Rb=2\sqrt{\Pb(1-\Pb)}\,,
\ee 
see Eqs.~\eqref{s1} and \eqref{rph}.
Using, e.g., scattering theory, it can be shown that
\be
\Av{\Psi^{\Rb}_0}{\hat{\mathcal{V}}} = 
\frac{4\pi}{\sqrt{\bar{Z}}}
\int d\epsilon\sum_{\alpha=-1}^1 f\left(\epsilon-\Phi\frac{\alpha}{2}\right)
\epsilon\,\rho^{\bar{Z}}_d(\epsilon)\,,
\label{analyticalAvV}
\ee
where
\be
\bar{Z}\equiv \Rb^2
\ee
can be interpreted as the quasiparticle weight of a single-particle 
excitation,~\cite{0953-8984-16-16-R01,Nozieres-Journal-of-Low-Temperature-Physics}
and
\bea
\rho_d^{\bar{Z}}(\epsilon) \equiv -\frac{1}{\pi}\text{Im}
\left[\frac{1}{\epsilon+i0^+ - \bar{Z}\Delta(\epsilon+i0^+)}\right]
\eea  
has the form of a renormalized uncorrelated impurity spectral function.

It is convenient to summarize here some of the Gutzwiller results 
derived in Ref.~\onlinecite{PhysRevB.82.195326} for the steady state and 
to underline the limits of the method from the quantitative point of view.
This will facilitate the subsequent analysis of the dynamics.

In the rest of this section we will assume the so called wide band limit,
i.e., $\Gamma<<W$.

At $\Phi=0$ and large $U/W$ one finds that 
\be
\bar{Z}\sim\frac{W}{\Gamma}\exp\left(-\frac{\pi}{16}\frac{U}{\Gamma}\right)
\equiv \frac{T_K^G}{\Gamma}\,,
\label{tkG}
\ee
The above expression for $T_K^G$ can be interpreted as the ``Gutzwiller approximation'' 
for the Kondo temperature. 

Notice that $T^G_K$ has two fundamental differences with respect to the correct value of 
the Kondo temperature~\cite{Hewson}
\be
T_K\sim U \sqrt{\frac{\Gamma}{2U}}
\exp\left(-\frac{\pi}{8}\frac{U}{\Gamma}+\frac{\pi}{2}\frac{\Gamma}{U}\right)\,:
\label{tkHew}
\ee
(i) the universal prefactor in the exponent should be $\pi/8$ and not $\pi/16$, 
and (ii) the factor $W/\Gamma$ (which is equal to $1/\Gamma$ in our units)
in Eq.~\eqref{tkG} diverges in the wide band limit.
The divergence of $T_K^G$ for $W\rightarrow\infty$
reflects the unreliability of the method in this limit.
In fact, it can be verified~\cite{PhysRevB.82.195326} that the Gutzwiller 
approximation predicts that $\bar{Z}\rightarrow 1$ when $W/\Gamma\rightarrow\infty$
independently on the value of $U$.

At finite $\Phi$ the steady-state quasi-particle weight $\bar{Z}$ vanishes 
for a finite $U=U_c$. 
This (unphysical) critical point occurs when 
\be
\Phi\sim T^G_K\,,\label{fakecrit}
\ee 
that is out of the regime of validity of the method.~\cite{PhysRevB.82.195326}

When the method is extended to out of equilibrium,
the inaccuracy in $\bar{Z}$ is reflected in the inaccuracy in the 
steady state current
\be
\bar{I}[\Psi] = -i\sum_{k\sigma} \frac{V_k}{\sqrt{\Omega}}\left(
\Av{\Psi}{\dc_\sigma\ca_{k\sigma,-1}} - \text{c.c.} \right)\,.
\ee
In fact, when $\ket{\Psi}$ is a Gutzwiller wavefunction of the form [Eq.~\eqref{GWFch4ph}]
the expectation value of the current in the steady state is given
by~\cite{PhysRevB.82.195326}
\be
\bar{I}[\Psi] = \int^{\frac{\Phi}{2}}_{-\frac{\Phi}{2}} d\epsilon\, 
\bar{Z}\Gamma(\epsilon)\,\rho_d^{\bar{Z}}(\epsilon)\,.
\ee

On the other hand, the method provides results in accordance with
the expected universality in terms of the Gutzwiller Kondo energy scale $T^G_K$ 
defined in Eq.~\eqref{tkG}.
In particular, it can be shown~\cite{PhysRevB.82.195326} that the 
approximated Gutzwiller conductance $G$ at zero bias is universal, and that 
its curvature at $\Phi=0$ has the expected $T^{G}_K$ dependence:
\be
\left.\frac{d^2G}{d\Phi^2}\right|_{\Phi=0}\sim \frac{-1}{(T^{G}_K)^2}\,.
\ee

\subsection{Interaction quench}\label{intquenchsection}

In this section we study the dynamics of the 
Anderson system [Eq.~\eqref{siam}] prepared in its nonequilibrium 
steady state at given bias $\Phi$
after a sudden change in $U$ at the initial time $t=0$. 
This way of perturbing the system is commonly referred to as 
an \emph{interaction quench}.

In order to assess the reliability of the variational ansatz 
defined in Eq.~\eqref{GWFch4ph} we focus on interaction quenches from $U=0$.
This is convenient as, for such a non-interacting system, the Gutzwiller 
steady-state formalism reproduces the exact solution.~\cite{PhysRevB.82.195326}
In our calculations we assume a flat density of states and 
$\Gamma/W=0.1$ (essentially the wide band limit).

\begin{figure}
  \centering
  \includegraphics{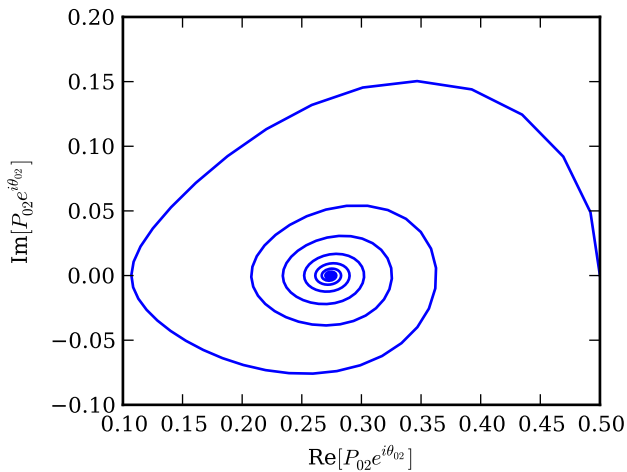}
  \caption{\label{fig:Dissipation}(Color online) Time evolution, for $\Gamma = 0.1$, 
    of the variational parameters $P_{02}$ and $\theta_{02}$ in an interaction quench
    from $U/\Gamma=0$ to $U/\Gamma = 4.0$, starting from the non-interacting steady 
    state with bias, $\Phi/\Gamma = 4.0$. $P_{02}$ starts from the non-interacting 
    value $P_{02}(t=0) = 1/2$ and evolves toward the interacting steady state value 
    $P_{02}(t\rightarrow \infty) \approx 0.27$.}
\end{figure}

When the interaction is turned on, a disturbance propagates away from the impurity 
through the leads with a speed that is given by the 
Fermi velocity of the leads $v_F$,~\cite{NewJPhys.12.043042}
and every local observable converge to its steady state value after some relaxation time
due to dissipation in the leads.~\cite{andrei-dissipation}
We remark that, unlike the case of the Hubbard model,~\cite{PhysRevLett.105.076401}
the Gutzwiller parameter $P_{02}$ --- which is the expectation value of the local 
operator $\hat{P}_{02}$, see Eq.~\eqref{p02def}, --- and its conjugate variable $\theta_{02}$
generally relax to proper steady state values, see Fig.~\ref{fig:Dissipation}.

It can be verified that the curvature of the current at $t=0$ 
obtained from Eq.~\eqref{tdGcurr} and Eqs.~\eqref{dyn2bisph}-\eqref{d2} is quantitatively 
correct at very short times, as
\bea
\frac{d^2I}{dt^2}(t=0)&=&-\frac{I(t=0)}{4}\,U^2\nonumber\\
&=&-\Av{\Psi_0}{[\hat{\h},[\hat{\h},\hat{I}]]}\,.
\eea
The fact that the approach provides quantitatively reliable results at short times 
is not surprising, as the Gutzwiller projector is expected to adequately describe the
physics of Coulomb blockade before the disturbance caused by the interaction 
quench has propagated away from the impurity, and other correlation effects 
has had time to build up and affect the dynamics of the system.

\begin{figure}
  \centering
  \includegraphics{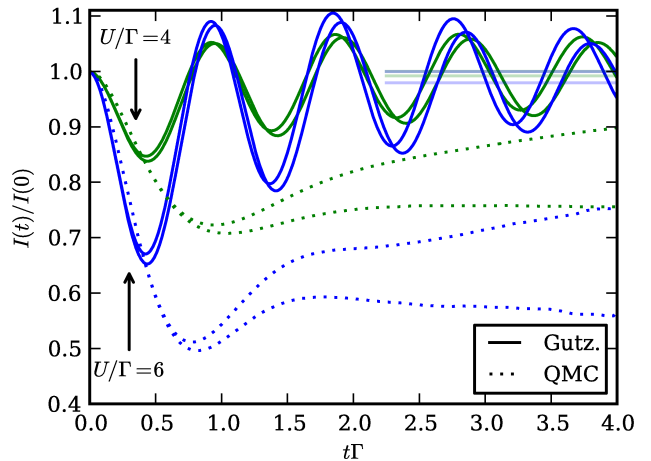}
  \caption{\label{fig:InteractionQuench_Bias_series}(Color online) Interaction quenches from $U/\Gamma = 0$ for $t<0$ to $U/\Gamma = 4$ and $6$ for $t>0$. 
    The scaled Gutzwiller transient current $I(t)/I(0)$ (solid lines) are shown for two biases $\Phi/\Gamma = 1/16$ and $3/4$ (higher bias giving higher $I(t\rightarrow \infty)$, indicated by horizontal lines). 
For comparison the corresponding quantum Monte Carlo results from Ref.~\onlinecite{diagrMC-werner2} are shown (dotted lines). 
}
\end{figure}

\begin{figure}
  \centering
  \includegraphics{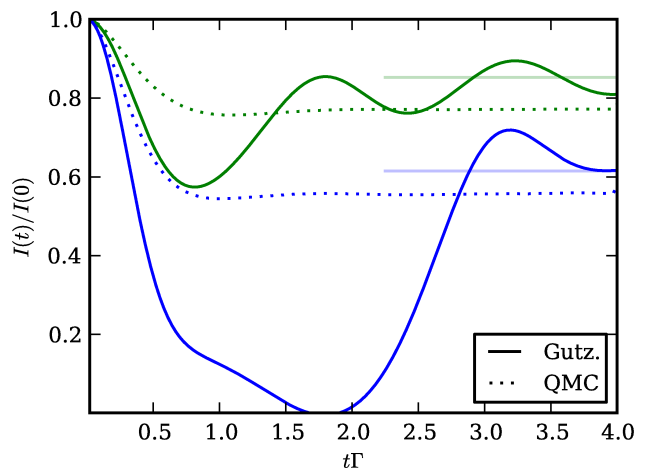}
  \caption{\label{fig:InteractionQuench_ShortTime}(Color online) Interaction quench from $U = 0$ to $U/\Gamma = 4$ and $6$ (counting from above) at a bias $\Phi/\Gamma = 4$ and $\Gamma = 0.1$ (solid lines). 
The corresponding steady-state currents are indicated by horizontal lines.
For comparison, quantum Monte Carlo results from Ref.~\onlinecite{diagrMC-werner2} for the same system (dotted lines).
  }
\end{figure}

In order to discuss the reliability of our approximation also at longer times
we have calculated the time evolution of the interacting current $I(t)$ divided by its value 
at $t=0$ for different biases and interaction strengths, see
Figs~\ref{fig:InteractionQuench_Bias_series}  and \ref{fig:InteractionQuench_ShortTime}.
Our calculations are compared with the numerically-exact Monte Carlo results of 
Ref.~\onlinecite{diagrMC-werner2} in two different regimes of parameters:
the weak bias regime $\Phi\lesssim T_K$,
and the large bias regime, $\Phi\gtrsim T_K$.

At $\Phi\lesssim T_K$, see Fig.~\ref{fig:InteractionQuench_Bias_series}, 
the quantum Monte Carlo normalized current $I(t)/I(0)$ is essentially
independent on the bias at short times, with a pronounced undershoot at 
$t\Gamma\sim 1$. 
The Gutzwiller time dependent current is in good agreement with the 
Monte Carlo results up to $t\Gamma\lesssim 0.4$.
At longer times, instead, the Gutzwiller current deviates from the correct solution. 
In particular, the double-occupancy oscillations in the impurity developed after the 
interaction quench are very weakly damped (see Sec.~\ref{ltnote}),
and induce unphysical oscillations also in the current. 
Notice that in the weak bias regime the magnetic correlations, that are not properly
taken into account in our approximation, play a very important role.
A Kondo cloud builds up and damp the charge fluctuations in the impurity 
induced by the interaction quench after a time $T$ given roughly by 
$T\lesssim 1/T_K$.
In Fig.~\ref{fig:InteractionQuench_Bias_series} this corresponds to
$T\Gamma\sim 2.3$ and $4.7$ for $U/\Gamma=4$ and  
$6$ respectively. Indeed, we interpret the pronounced minimum of the current at 
$t\Gamma\sim 1$ and the subsequent small oscillations visible at 
longer times for $U/\Gamma=6$ in the Monte Carlo calculations
as a consequence of oscillations in the double occupancy.  
A better variational description of the magnetic correlations between 
the impurity and the surrounding lead electrons is necessary to reproduce 
this damping.

At $\Phi\gtrsim T_K$, see Fig.~\ref{fig:InteractionQuench_ShortTime},
the the quantum Monte Carlo current depends on the bias even at short times.
The time $T$ required for convergence to the steady state
is much shorter than in the small bias regime, and is essentially 
independent of $U$.~\cite{diagrMC-werner2}
In this case the Gutzwiller time dependent current is in good 
agreement with the Monte Carlo calculations only at very short times,
$t\Gamma\lesssim 0.1$. For longer times
the damping rate of the current oscillations is underestimated as 
in the weak bias regime.
Note that the influence of nonequilibrium effects around the impurity 
is strong in this regime. 
Indeed, it is known that the bias can destroy Kondo effect 
when $\Phi\gtrsim T_K$.~\cite{kondo-destruction}
We believe that the fast damping of the oscillations of the quantum
Monte Carlo current is a manifestation of this nonequilibrium 
effect, which is sufficiently strong to 
contrast the formation of a magnetic moment already at $t\Gamma\ll 1$. 
Although the destruction of the Kondo resonance can be qualitatively 
described already in our approximation,~\cite{PhysRevB.82.195326} 
a reliable description of the quantum fluctuations
induced by the bias in the vicinity of the impurity would require an
extended Gutzwiller projector.

In conclusion, the Gutzwiller dynamics is quantitatively accurate at sufficiently short times
(depending on the regime of parameters considered).
At longer times, instead, the Gutzwiller dynamics is no more reliable.
In particular, the charge oscillations developed by the current after the 
interaction quench are very weakly damped.

\subsection{Bias quench}

In this section we study the dynamics of the system prepared in a given equilibrium 
configuration after a sudden shift of the chemical potentials $\mu_\alpha$ in the leads; 
a so called \emph{bias quench}.
We limit the discussion to symmetric quenches from zero to finite chemical potentials
$\mu_\alpha = \alpha\Phi/2$.

In our calculation we adopt the following ``smoothed'' square density of states in the leads
\begin{align}
  \rho(\epsilon) = \frac{1}{2}
   \frac{1}{1 + e^{\nu ( \epsilon - W)}}
   \frac{1}{1 + e^{-\nu ( \epsilon + W)}} 
\end{align}
with $\nu \Gamma = 3$, where $\nu$ is a measure of smoothness of the edges of the DOS and $W=1$ 
is the half-bandwidth.

\begin{figure}
  \centering
  \includegraphics{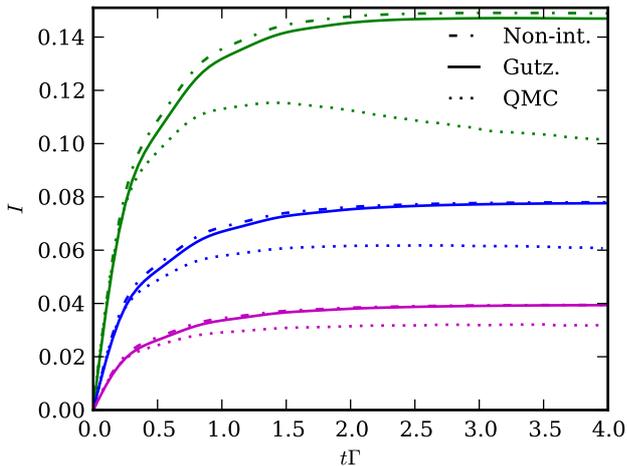}
  \caption{\label{fig:BiasQuench}(Color online) Current $I$ for bias quenches from the unbiased state $\Phi/\Gamma = 0$ to finite bias $\Phi/\Gamma = 0.25$, $0.5$ and $1.0$ (higher bias correspond to higher current) for $U/\Gamma = 4$. The Gutzwiller results (solid lines) are compared to the non-interacting quench (dash dotted lines) and quantum Monte Carlo results from Ref.~\onlinecite{diagrMC-werner2} (dotted lines).
}
\end{figure}

\begin{figure}
  \centering
  \includegraphics{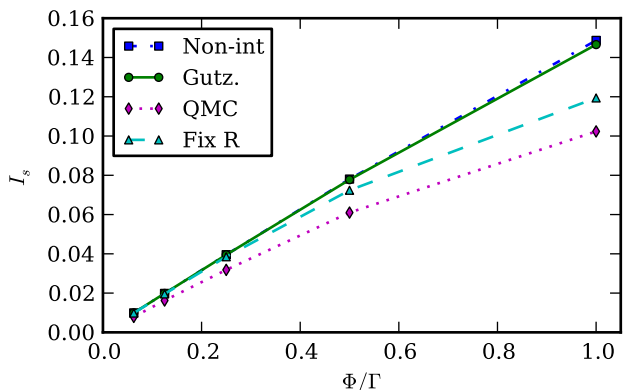}
  \caption{\label{fig:SteadyStateCurrent}(Color online) Steady state currents $I_s$
    in the weak bias regime. The Gutzwiller results (circles) are compared to the 
    non-interacting bias quench (squares) and quantum Monte Carlo results (diamonds) 
    from Ref.~\onlinecite{diagrMC-werner2}. The effect of using the 
    ``a priori'' estimate of the Kondo temperature to fix $R$ is also shown (triangles).}
\end{figure}

In our time dependent calculations the system is initially prepared
in its approximated interacting Gutzwiller equilibrium state.
Unfortunately, unlike the case of the interaction quench, the initial state can not be prepared
exactly in the Gutzwiller approximation, as $U$ is finite even at $t=0$.
For this reason we can not attempt a quantitative verification of the Gutzwiller dynamics
in this case, not even at short times.
The transient current after the application of weak biases $\Phi\le\Gamma$ is shown in 
Fig.~\ref{fig:BiasQuench} for three different interaction strengths, $U/\Gamma = 0, 4$ and $6$. 
In this weak bias regime the Gutzwiller parameters $P_{02}$ and $\theta_{02}$ 
do not change appreciably after the bias quench, independently of $\Gamma$ and $U$.
Consequently, the current evolves in time essentially as in a renormalized uncorrelated system
$\hat{\h}^R_0$, see Eq.~\eqref{hrenph}, with a constant renormalization $R$ 
giving an overall reduction in the current at finite $U$. 
In the special case of $U=0$ the renormalization constant $R$ is equal to $1$,
and the Gutzwiller calculation  reproduces the exact dynamics of the non-interacting 
system.

The shoulder like features in the transient current in Fig.~\ref{fig:BiasQuench}, 
that are present even in absence of interaction, are due to resonance effects induced 
by the specific form of the density of states,~\cite{GokerNCA}
and would not be visible in the wide band limit.

From the comparison of our calculations with the quantum Monte Carlo
results of Ref.~\onlinecite{diagrMC-werner2} it is clear that the suppression 
of the current due to $U$ is severely underestimated in the Gutzwiller approximation.
This is due to the problem discussed in Sec.~\ref{steadystatesection} 
that $R\sim 1$ for large bandwidths, even when $U>\Gamma$, causing the
Gutzwiller dynamics to be essentially generated by $\hat{\h}_0$ for weak bias quenches.

Although a quantitative variational description of the dynamics is impossible 
with the oversimplified ansatz [Eq.~\eqref{GWFch4ph}], the fact that 
the Gutzwiller current behaves essentially as for a renormalized uncorrelated system $\hat{\h}^R_0$
for small bias quenches suggests a simple interpretation of some of the qualitative features of the 
current provided by the Monte Carlo calculations.
In Fig.~\ref{fig:SteadyStateCurrent} is shown the steady state current generated after a bias 
quench by $\hat{\h}^R_0$, with
\be
R=\sqrt{T_K/\Gamma}\,,
\ee
where $T_K$ is given by Eq.~\eqref{tkHew} rather than the Gutzwiller approximation.
This simple calculation better reproduces the steady state current
of the quantum Monte Carlo calculations.
This supports our statement that the main problem of the time dependent 
Gutzwiller method for small bias quenches is that the renormalization factor $R$ is  
underestimated in the initial equilibrium state.

\subsection{Note about relaxation and long-time behavior}\label{ltnote} 

\begin{figure}
  \centering
  \includegraphics{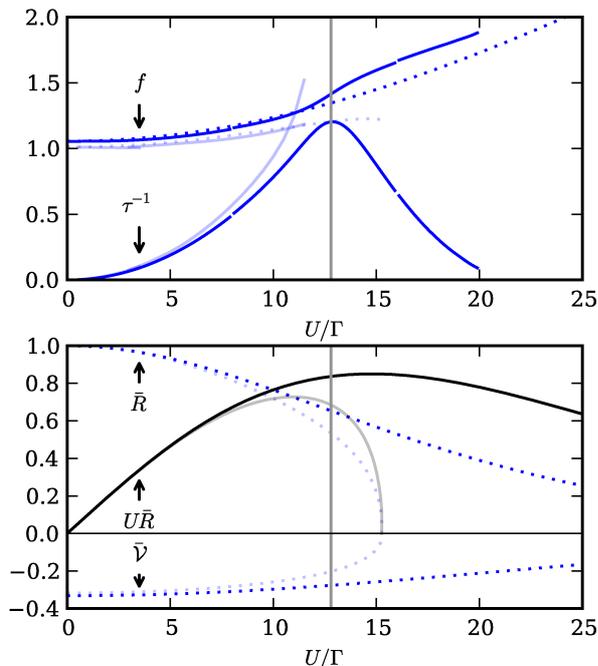}
  \caption{\label{fig:FreqDamp}(Color online) Upper panel. Oscillation frequency $f$ and inverse relaxation time $1/\tau$ as a function of $U\Gamma$ for zero and finite bias,  $\Phi/\Gamma = 0$ and $1$ (full and shaded lines respectively). 
The dotted line represents the frequency $f_0$ defined in Eq.~\eqref{fr}.
Lower panel. Steady state values of $\bar{R}$ and $\bar{\mathcal{V}}$. 
The maximum of $\tau^{-1}$ at $\Phi=0$, that is realized at $U\Gamma \approx 12.8$, is indicated by the vertical line.}
\end{figure}

As we have shown, the Gutzwiller dynamics is 
able to capture dissipation in the leads, and a Gutzwiller steady-state 
can generally be reached after a sufficient thermalization time. 
In this section we discuss the long-time dynamics more in detail, 
and we analyze its dependency on the tight binding parameters
of the Hamiltonian and the bias.

In Sec.~\ref{steadystatesection} 
we anticipated that at finite bias the steady-state theory is qualitatively reliable only 
for $U\ll U_c$, where $U_c$ is the spurious critical interaction
discussed in Sec.~\ref{steadystatesection},
characterized by a vanishing quasi-particle weight $Z=R^2$.
For this reason we confine our analysis to $U\ll U_c$, which corresponds to
the weak bias regime defined by the condition
\be
\Phi\ll T^G_K\,,\label{regan}
\ee
see Eq.~\eqref{fakecrit}.

Within the range of parameters considered, we find that the long-time behavior of $\P(t)$ 
is, to a good approximation, an exponentially damped oscillation with a measurable
frequency $f$ and a relaxation time $\tau$.

Interestingly, we find that both $\tau$ and $f$ are independent
on the initial condition, and depend solely on 
the bias and the tight binding parameters of the Hamiltonian (not shown).
The dependency of $\tau$ and $f$ on $U$ is illustrated in the upper panel of
Fig.~\ref{fig:FreqDamp} for zero and finite bias.

Our numerical results show that the Gutzwiller dynamics spontaneously reaches 
a steady state except in
the following
three limiting cases: 
(i) at $U=0$, 
(ii) at $U\rightarrow\infty$ and $\Phi=0$, and 
(iii) for $U\geq U_c$ at finite bias.
The inverse relaxation time $\tau^{-1}$ increases by increasing $U$ from $0$ 
up to a maximum value, and vanishes afterwards.
Note that the effect of the bias is to decrease the relaxation time for any given value of $U$.
The frequency of the oscillations $f$ grows monotonically with $U$,
and is smaller at finite bias with respect to the equilibrium case.

In order to better discuss our numerical results we
linearize Eqs.~\eqref{dyn1bisph}-\eqref{d2} around the steady state solution.~\cite{schironew}
We introduce steady state values (indicated by bars) and deviations from the steady state
\bea
  \P &=& \Pb + \Pd \nonumber \\
  \t &=& \tb + \td \nonumber \\
  \V &=& \Vb + \Vd\,.
\eea
It can be easily shown that
\bea
  \partial_t \Pd &=& \Rb \Vb \td \nonumber \\
  \partial_t \td &=& - \frac{U}{2\Vb} \Vd - \frac{4\Vb}{\Rb^3} \Pd\,,
\eea

Eliminating $\td$ we obtain the following equation for $\Pd$:
\be
\partial^2_t \Pd = - \frac{U\Rb}{2} \Vd - \frac{4\Vb^2}{\Rb^2} \Pd\,.
\label{ltb}
\ee

Notice that if $U\Rb=0$ Eq.~\eqref{ltb} reduces to the 
differential equation of a simple harmonic oscillator for $\Pd$, 
with a resonance frequency 
\be
f_0 \equiv \frac{1}{2\pi} \frac{2|\Vb|}{\Rb}\,.\label{fr}
\ee
This observation explains why the expectation value of the double occupancy does not 
relax whenever either $U$ or $\Rb$ is zero. 
In particular, Eq.~\eqref{ltb} shows that the critical point at finite bias, 
see Eq.~\eqref{tkG},  is reflected by a spurious
dynamical transition, characterized by a diverging relaxation time.

The result that the relaxation time of the system diverges
in the large $U$ limit at $\Phi=0$ is physically correct, as in this limit the only
energy scale is the Kondo temperature, which vanishes exponentially with $U$, 
see Eq.~\eqref{tkHew}.
Also the fact that the relaxation time $\tau$ decreases by increasing the 
bias~\cite{diagrMC-werner2}, which reflects the fact that bias 
``destroys'' Kondo effect,~\cite{kondo-destruction}
is qualitatively captured by the Gutzwiller approximation.
On the contrary, the fact that the expectation value of the double occupancy 
is unable to relax at $U=0$ is clearly a drawback of our oversimplified variational
ansatz.~\cite{andrei-dissipation}
This incorrect feature of the Gutzwiller solution questions 
the whole observed behavior of $\tau^{-1}$
for $U\Gamma\lesssim 12.8$ in Fig.~\ref{fig:FreqDamp}.

Let us now consider the behavior of the frequency $f$.
From Fig.~\ref{fig:FreqDamp} we see that $f \sim f_0$ for all values of $U$. 
Using Eq.~\eqref{analyticalAvV},
it can be readily shown that
\be
\lim_{U\rightarrow\infty}
\left(
\frac{1}{2\pi} \frac{2|\Vb|}{\Rb}
\right)
=\infty\,.\label{limfrq}
\ee
This limit exposes another drawback of the approximation, as the
same dimensional argument used above for the relaxation time should be
applicable to the period of oscillation:
at large $U$ the only time scale is the inverse of the Kondo temperature, 
which vanishes exponentially for $U\rightarrow\infty$.
For this reason we conclude that the behavior of $f$ can not be addressed 
within our variational approximation.

In conclusion, the simple variational ansatz [Eq.~\eqref{GWFch4ph}] is 
able to capture dissipation in the leads,
but a more general type of variational wavefunction is necessary in order to obtain
a reliable description of the long time behavior.

\section{Conclusions}\label{conclusions} 

We have generalized the time dependent Gutzwiller approach of 
Ref.~\onlinecite{PhysRevLett.105.076401} to impurity problems, 
modeling a quantum system, e.g., a molecule or a quantum dot, coupled to metallic leads.
The dissipation effects in the junction due to the coupling to the leads
are naturally accounted for by the method, and a steady state is reached spontaneously, 
without, e.g., including any fictitious bosonic bath.~\cite{MuhlbacherandRabani}

We have shown that the time dependent theory, that is based~\cite{PhysRevB.83.165105} 
on the Dirac-Frenkel variational principle,~\cite{dirac,frenkel}
enables a natural derivation of a variational 
principle for the steady state, that reproduces the
ansatz previously proposed on different grounds in Ref.~\onlinecite{PhysRevB.82.195326}.
We believe that our formulation of the steady state problem represents
an important conceptual advancement in itself, and that the basic idea behind
it could be exploited also to develop different numerical techniques.

The Gutzwiller method for the steady state is particularly appealing from the 
computational point of view as, although there are important conceptual differences between 
the Gutzwiller theory for the nonequilibrium steady state and the conventional method
at equilibrium, it seems to us that these differences are not accompanied by any 
substantial increase in computational complexity --- which is generally the case 
for other methods.

We have investigated the reliability of the time dependent variational method in the case
of a Gutzwiller projector acting only on the impurity, see Eq.~\eqref{GWFch4ph}, 
which is the simplest conceivable type of Gutzwiller variational function.
From the comparison with the quantum Monte Carlo data of Ref.~\onlinecite{diagrMC-werner2} 
we concluded that the obtained Gutzwiller dynamics 
after an interaction quench is quantitatively accurate at sufficiently short times.
Nevertheless, due to the limited description of the quantum fluctuations and magnetic correlations
between the impurity and the surrounding electrons, the damping rate of the 
current oscillations generated in the impurity by the sudden rise of the 
interaction $U$ at $t=0$ is underestimated in this approximation, 
and the steady state current is quantitatively incorrect.

Considering the fact that the disturbance due to the interaction quench propagates with
a finite velocity $v\sim v_F$ from the impurity, the behavior of the current
will be better described, at least at short times, 
by extending the region of action of the Gutzwiller projector
to a portion of the leads surrounding the impurity. 
But we also expect that this more elaborate variational ansatz would provide 
a better description of the time dependent current at any time, uniformly. 
In fact, the correction due to the Gutzwiller 
projector on the correlation functions is expected to vanish 
at distances sufficiently large from the impurity due to the screening of the 
Coulomb interaction provided by the conduction electrons in the vicinity of the impurity.
The numerical verification of this hypothesis will constitute an interesting 
extension of this work.

\begin{acknowledgments}

It is a pleasure to thank Giovanni Morchio and Michele Fabrizio 
for insightful discussions that allowed us to clarify several important 
points related to this work.
We also thank Bo Hellsing, Natan Andrei, Philipp Werner and Dirk Schuricht 
for constructive discussions. 
\end{acknowledgments}

\bibliographystyle{apsrev}


\end{document}